\documentclass[copyright,creativecommons]{eptcs}
\providecommand{\event}{E. Formenti (Ed.): AUTOMATA \& JAC 2012} 

\usepackage[usenames]{color}
\usepackage{qtree}
\usepackage[dvips]{graphicx}
\usepackage{gastex}
\usepackage{amsmath}
\usepackage{amssymb}
\usepackage{verbatim}
\usepackage[LY1]{fontenc}
\usepackage{amssymb}
\usepackage{enumerate}
\usepackage{amsfonts}
\usepackage[all]{xy}
\usepackage{amscd}
\usepackage{latexsym}
\usepackage{amsthm}
\usepackage{amsbsy}
\usepackage{subfigure}
\usepackage{eufrak}

\numberwithin{equation}{section}

\newcommand {\N}{\mathcal{N}} 
\newcommand{\F}{\mathcal{F}}
\newcommand{\M}{\mathcal{M}}

\newcommand{\B}{\mathcal{B}}

\newcommand{\C}{\mathcal{C}}

\DeclareMathOperator{\dist}{dist}

\DeclareMathOperator{\Sh}{\mathsf{X}}

\DeclareMathOperator{\supp}{supp}

\theoremstyle{plain}
\newtheorem{theorem}{Theorem}

 \newtheorem{lemma}[theorem]{Lemma}
 \newtheorem{proposition}[theorem]{Proposition}
 
 \newtheorem{corollary}[theorem]{Corollary}
\theoremstyle{definition}
 \newtheorem{definition}[theorem]{Definition}

\theoremstyle{remark}
 \newtheorem{remark}[theorem]{Remark}
 \newtheorem{example}[theorem]{Example}
\theoremstyle{remark}

\makeatletter
\def\vneed#1{\par \penalty-100 \begingroup 
   \ifdim\prevdepth>\maxdepth \dimen@i\maxdepth \else \dimen@i\prevdepth\fi
   \kern-\dimen@i
   \dimen@\pagegoal \advance\dimen@-\pagetotal 
   \ifdim #1>\dimen@ \vfil \eject \fi
   \kern\dimen@i
   \endgroup}
\makeatother

\title{Topological properties of cellular automata on trees}

\author{Gabriele Fici
\institute{Laboratoire I3S\\ CNRS \& Universit\'e Nice Sophia Antipolis\\ 06903 Sophia Antipolis, France}
\email{fici@i3s.unice.fr}
\and
Francesca Fiorenzi
\institute{Laboratoire de Recherche en Informatique\\
Universit\'e Paris-Sud 11\\
91405 Orsay Cedex, France}
\email{\quad fiorenzi@lri.fr}
}

\def\titlerunning{Topological properties of cellular automata on trees}
\def\authorrunning{G. Fici and F. Fiorenzi}
\begin{document}
\maketitle

\begin{abstract}
We prove that there do not exist positively expansive cellular automata defined on the full $k$-ary tree shift (for $k \geq2$).
Moreover, we investigate some topological properties of these automata and their relationships, namely permutivity, surjectivity, preinjectivity, right-closingness and openness.\\


\end{abstract}

\section{Introduction}
In the classical theory of cellular automata (CA), the universe is the grid $\mathbb{Z}^d$ of integer points of the Euclidean $d$-dimensional space. The state of every cell in the grid ranges in a finite nonempty set $A$.
A configuration is an element of $A^{\mathbb{Z}^d}$, that is, a map $f \colon \mathbb{Z}^d \to A$ that describes the state of every cell. A cellular automaton is a map $\tau\colon A^{\mathbb{Z}^d}\to A^{\mathbb{Z}^d}$ that changes a configuration by simultaneously updating the state of each cell according to a fixed local rule, i.e., a rule that only considers the states of the neighbors of this cell.

In this setting, an important role is played by one-dimensional CA ($d=1$) and many results holding in this case no longer hold in the multi-dimensional case ($d\geq2$). Another framework that has been extensively studied is that of (one-dimensional) one-sided CA (i.e., those defined on $A^\mathbb{N}$).

Provided that the universe remains discrete and homogeneous, CA can be considered in a much more general setting, in which the grid is replaced by the Cayley graph of a finitely generated group or semigroup $G$. Notice that the grid $\mathbb{Z}^d$ is the Cayley graph of the free abelian group of rank $d$. It is well known that $A^G$ is a compact metric space, and there is a natural right action of $G$ on $A^G$, called the shift action. CA are characterized as the continuous self-mappings of $A^G$ that commute with this action.

\medskip

In this paper we study topological properties of CA on $A^{\Sigma^*}$, where $\Sigma^*$ is the free monoid of finite rank $\vert\Sigma\vert$. This generalizes the case of one-sided CA (where $\vert\Sigma\vert=1$). The Cayley graph of $\Sigma^*$ is a regular $\vert\Sigma\vert$-ary rooted tree.
This setting was recently studied in~\cite{Aubrun11, AubrunBeal10, AubrunBeal12,CeccheriniCoonaertFiorenziSunic12}. Notice that topological properties of CA acting on configurations spaces different from
$A^{\mathbb{Z}^d}$ have already been studied, namely in the framework of sand automata (see~\cite{DennunzioGuillonMasson09}).

A property that holds only in the one-dimensional case concerns expansivity. Indeed, there do not exist multidimensional CA that are positively expansive~\cite{Shereshevsky93}, whilst one-dimensional CA and also one-sided CA can have this property~\cite{BlanchardMaass95, BlanchardMaass97}. We prove that there do not exist CA on $A^{\Sigma^*}$ that are positively expansive for $\vert\Sigma\vert\geq2$.

For one-dimensional CA (including the one-sided case), one can define left-permutivity and right-permutivity. We extend the definition of left-permutivity to CA on $A^{\Sigma^*}$ and we refer to this property as permutivity. Actually, due to the non-linearity of $\Sigma^*$, the definition of right-permutivity cannot be naturally extended. We prove that permutive CA are surjective and preinjective. Despite the fact that the surjectivity and the preinjectivity of a CA on $A^{\mathbb{Z}^d}$ are equivalent in any dimension $d$,  we show an example of a surjective CA on $A^{\Sigma^*}$ which is not preinjective. On the other hand, we prove that preinjectivity implies surjectivity.

We extend the notion of right-clonsingness of a CA in our setting, and we prove that it implies preinjectivity.

In~\cite{BlanchardMaass97}, Blanchard and Maass prove that right-closing one-sided CA are open. We extend their proof to our class of CA of radius one, but we believe that the result holds true in general.

At the end of the paper we provide 
some concluding remarks and we briefly discuss the perspectives of future work.

\subsection*{Acknowledgements}
We thank the referees for their helpful comments. We are especially grateful to the referee who found out a logical mistake in the proof of Lemma~\ref{l:existence t} and suggested the correct statement of Proposition~\ref{p:card=}. This in turn allowed us to significantly improve the results presented in this paper.

\section{Definitions and background}

\subsection{The rooted tree $\Sigma^*$}\label{ss:free}
For a positive integer $k$, we denote by $\Sigma$ the set $\{0,1, \dots,k-1\}$.
Let $n \in \mathbb{N}$ be a nonnegative integer. We denote by $\Sigma^n$ the set of all \emph{words} $v = \sigma_1\sigma_2 \cdots \sigma_n$ of \emph{length} $n$ (where $\sigma_i \in \Sigma$ for $i=1,2,\dots,n$) over $\Sigma$. In particular, $\varepsilon \in \Sigma^0$ indicates the only word of length $0$, called the \emph{empty word}. For $n \geq 1$, we denote by $\Delta_n$ the set $\bigcup_{i = 0}^{n-1}\Sigma^i$ (that is, the set of all words of length $\leq n-1$). Notice that $\vert\Delta_n\vert = \frac{k^n-1}{k-1}$.

The \emph{concatenation} of two words $v = \sigma_1\sigma_2 \cdots \sigma_n \in \Sigma^n$ and $v' = \sigma'_1\sigma'_2 \cdots \sigma'_m\in \Sigma^m$ is the word $vv' = \sigma_1\sigma_2 \cdots \sigma_n\sigma'_1\sigma'_2 \cdots \sigma'_m\in \Sigma^{m+n}$. Then the set $\Sigma^* = \bigcup_{n \in \mathbb{N}} \Sigma^n$, equipped with the multiplication given by concatenation, is a monoid whose identity element is $\varepsilon$. It is called the \emph{free monoid} over the set $\Sigma$.

From the graph theoretical point of view, we consider $\Sigma^*$ as the vertex set of the regular $k$-ary rooted tree, where $k = \vert \Sigma\vert$. The empty word $\varepsilon$ is its root, and for every vertex $v \in \Sigma^*$ the vertices $v\sigma\in \Sigma^*$ (with $\sigma \in \Sigma$) are called the \emph{children} of $v$. Every vertex is connected to each of its children by a non-labeled edge.

\subsection{Configurations and shift spaces}
Let $A$ be a nonempty finite set, called the \emph{alphabet}. The elements of $A$ are called \emph{letters}.
The \emph{space of configurations} of $\Sigma^*$ over the alphabet $A$ is the set $A^{\Sigma^*}$ of
all maps $f \colon  \Sigma^* \to  A$. When equipped with the \emph{prodiscrete topology} (that is, with the product topology  where each factor $A$ of $A^{\Sigma^*} = \prod_{v \in \Sigma^*}A$ is endowed with the discrete topology) the configuration space is a compact, totally disconnected, metrizable space.
Also, the free monoid $\Sigma^*$ has a right action on $A^{\Sigma^*}$ defined as follows: for every $v \in \Sigma^*$ and $f \in  A^{\Sigma^*}$ the configuration $f^v \in  A^{\Sigma^*}$ is defined by setting
$$f^v(v') = f(vv'),$$ for all $v' \in \Sigma^*$. This action, called the \emph{shift action}, is continuous with respect to the prodiscrete topology.


Recall that a sub-basis for the prodiscrete topology on $A^{\Sigma^*}$ consists of the \emph{elementary cylinders}
$$\C(v,a) = \{f \in A^{\Sigma^*} : f(v) = a\},$$
where $v \in \Sigma^*$ and $a \in A$. In what follows, $\C(a)$ is an abbreviation for $\C(\varepsilon,a)$.
A \emph{cylinder} is a finite intersection of elementary cylinders. If $M \subset \Sigma^*$ is finite and $p \colon M \to A$ is a map, we denote by $\C(p)$ the cylinder
determined by $p$.

A neighborhood basis of a configuration $f \in A^{\Sigma^*}$ is given by the sets
$$\N(f,n) = \{g \in A^{\Sigma^*} : g\vert_{\Delta_n} = f\vert_{\Delta_n}\},$$
where $n \geq 1$ (as usual, for $M\subset \Sigma^*$, we denote by $f\vert_M$ the restriction of $f$ to $M$).

If
$f_1, f_2 \in A^{\Sigma^*}$ are two different configurations, we define
the distance
$$\dist(f_1,f_2) = \frac 1 n,$$
where $n = \min\{n \in \mathbb{N} : f_1 \neq f_2 \textup{ on } \Delta_n\}$. If $f_1 = f_2$, we set
their distance equal to zero. Notice that the topology induced by
this metric is just the product topology.

\begin{definition}[\bf Subshift]
A subset $X \subset A^{\Sigma^*}$ is called a \emph{subshift} (or \emph{tree shift}, or simply \emph{shift}) provided that $X$ is closed (with
respect to the prodiscrete topology) and \emph{shift-invariant} (that is, $f^v \in X$ for all
$f \in X$ and $v \in \Sigma^*$). In particular $A^{\Sigma^*}$ is a tree shift and it is called the \emph{full (tree) shift}.
\end{definition}

\subsection{Forbidden blocks and shifts of finite type}

\begin{definition}[\bf Pattern and block]
Let $M\subset \Sigma^*$ be a finite set. A \emph{pattern} is a map $p \colon  M \to  A$. The set $M$ is called the \emph{support} of $p$ and it is denoted by ${\rm supp}(p)$. We denote by $A^M$ the set of all patterns with support $M$. For any $n\geq1$,
a \emph{block} is a pattern $p \colon  \Delta_n \to  A$. The integer $n$ is called the \emph{size} of the block.
The set of all blocks is denoted by $\B(A^{\Sigma^*})$.
\end{definition}


If $X$ is a subset of $A^{\Sigma^*}$ and $M \subset \Sigma^*$ is finite,
the set of patterns $\{f\vert_M : f \in X\}$ is denoted by $X_M$. For $n \geq 1$, the notation $X_n$ is an abbreviation for $X_{\Delta_n}$ (that is, the set of all blocks of size $n$ which are restrictions to $\Delta_n$ of some configuration in $X$). We denote by $\B(X)$ the set of all blocks of $X$ (that is, $\B(X)= \bigcup_{n\geq1}X_n$).

Given a block $p \in \B(A^{\Sigma^*})$ and a configuration $f \in A^{\Sigma^*}$, we say that $p$ \emph{appears} in $f$ if there exists $v \in \Sigma^*$ such that $(f^v)\vert_{{\rm supp}(p)} = p$. If $p$ does not appear in $f$, we say that $f$ \emph{avoids} $p$.
Let $\F$ be a set of blocks. We denote by $\mathsf{X}(\F)$ the set of configurations in $A^{\Sigma^*}$ avoiding simultaneously all the blocks in $\F$, in symbols
$$\mathsf{X}(\F) = \{f \in A^{\Sigma^*} : (f^v)\vert_{\Delta_n} \notin \F, \textup{ for all } v \in \Sigma^* \textup{ and } n \geq 1 \}.$$

If $\vert \Sigma \vert = 1$ we have a one-dimensional setting in which $\Sigma^*$ is identified with $\mathbb{N}$. Indeed, if $\Sigma = \{0\}$, we associate $n\in \mathbb{N}$ with $0^n\in \Sigma^n$, where $0^n$ denotes the word $\underbrace{0\cdots 0}_n$. In this case, a configuration $f \in A^\mathbb{N}$ can be identified with the (right) infinite word $w = a_0a_1\cdots$ over the alphabet $A$ where $a_0 = f(\varepsilon)$ and $a_n = f(0^n)$ for all $n\geq 1$. Analogously, a block of size $n$ can be identified with an element of $A^n$, that is a word of length $n$ over the alphabet $A$. Indeed the set $\Delta_n = \{\varepsilon, 0, 00, \dots, 0^{n-1}\} \subset \Sigma^*$ is identified with $\{0, 1, 2, \dots, n-1\} \subset \mathbb{N}$.

By analogy with the one-dimensional case (see for example \cite[Theorem 6.1.21]{LindMarcus95}), we have the following combinatorial characterization of subshifts.

\begin{proposition}\label{p:forbidden}
A subset $X  \subset A^{\Sigma^*}$ is a subshift if and only if there exists a set
$\F \subset \B(A^{\Sigma^*})$ of blocks such that
$X = \mathsf{X}(\F)$.
\end{proposition}

Let $X \subset A^{\Sigma^*}$ be a subshift. A set $\F$ of blocks as in Proposition~\ref{p:forbidden} is called a \emph{defining set of forbidden blocks} for $X$.
If one can find a finite defining set of forbidden blocks for $X$, the subshift $X$ is called \emph{of finite type}.

\begin{remark}\label{r:B(X) determines X}The blocks of a subshift determine the subshift. Indeed, given two subshifts $X,Y \subset A^{\Sigma^*}$, we have
$X = \Sh(\B(A^{\Sigma^*}) \setminus \B(X))$,
so that
$X = Y \Longleftrightarrow \B(X) = \B(Y).$
\end{remark}

\subsection{Cellular automata}
\begin{definition}[\bf Cellular automaton]
A map $\tau \colon   A^{\Sigma^*} \to  A^{\Sigma^*}$ is called a \emph{cellular automaton} (CA for short) if there exist an integer $r\geq1$ and a map $\mu \colon  A^{\Delta_{r+1}} \to  A$ such that $$\tau(f)(v) = \mu((f^v)\vert_{\Delta_{r+1}})$$ for all $f \in A^{\Sigma^*}$ and $v \in \Sigma^*$. The integer $r$ is called the \emph{radius} of $\tau$, and $\mu$ is the associated \emph{local defining map}.
\end{definition}

The following is a topological characterization of cellular automata. For a proof in the one-dimensional case, see \cite[Theorem 6.2.9]{LindMarcus95}. See also \cite[Theorem 1.8.1]{livre} and \cite[Proposition 1.2.4]{Fiorenzi00thesis}, for a more general setting.

\begin{theorem}[\bf Curtis-Hedlund-Lyndon]\label{t:curtis}
A map $\tau \colon A^{\Sigma^*} \to  A^{\Sigma^*}$ is a cellular automaton if and only if it is continuous (with respect to the prodiscrete topology) and commutes with the shift action (that is, $(\tau(f))^v = \tau(f^v)$ for all $f \in A^{\Sigma^*}$ and $v \in \Sigma^*$).
\end{theorem}

\begin{example}
Let $A = \{0,1\}$ and $r=1$. The local defining map $\mu \colon A^{\Delta_2} \to A$ is defined by
\[
\begin{split}
\mu(p) & =
\begin{cases}
0 & \mbox{if } p(\epsilon)=p(0)=p(1)=0\\
1 & \mbox{otherwise}
\end{cases}
\end{split}
\]
for each $p \in A^{\Delta_2}$. In other words, $\mu$ sends a pattern $p\in A^{\Delta_{2}}$ in $0$ if and only if all the elements of the pattern are $0$'s. An illustration of the behavior of the CA $\tau$ defined by $\mu$ is depicted in Figure \ref{fig:fici}.

\begin{figure}[!h]
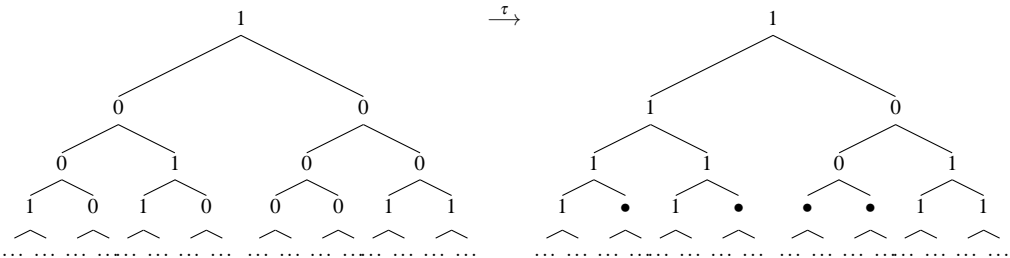

\begin{center}
\qtreecenterfalse
\scriptsize
\Tree [.$1$ [.$0$ [.$0$ [.$1$ $\dots$ !\qsetw{0.1cm} $\dots$ ] !\qsetw{0.5cm} [.$0$ $\dots$ !\qsetw{0.1cm} $\dots$ ] ] !\qsetw{1cm} [.$1$ [.$1$ $\dots$ !\qsetw{0.1cm} $\dots$ ] !\qsetw{0.5cm} [.$0$ $\dots$ !\qsetw{0.1cm} $\dots$ ] ] ] !\qsetw{3cm} [.$0$ [.$0$ [.$0$ $\dots$ !\qsetw{0.1cm} $\dots$ ] !\qsetw{0.5cm} [.$0$ $\dots$ !\qsetw{0.1cm} $\dots$ ] ] !\qsetw{1cm} [.$0$ [.$1$ $\dots$ !\qsetw{0.1cm} $\dots$ ] !\qsetw{0.5cm} [.$1$ $\dots$ !\qsetw{0.1cm} $\dots$ ] ] ] ]
\hspace{0.1cm}$\stackrel{\tau}{\longrightarrow}$\hspace{0.1cm}
\Tree [.$1$
[.$1$ [.$1$ [.$1$ $\dots$ !\qsetw{0.1cm} $\dots$ ] !\qsetw{0.5cm} [.$\bullet$ $\dots$ !\qsetw{0.1cm} $\dots$ ] ] !\qsetw{1cm} [.$1$ [.$1$ $\dots$ !\qsetw{0.1cm} $\dots$ ] !\qsetw{0.5cm} [.$\bullet$ $\dots$ !\qsetw{0.1cm} $\dots$ ] ] ]
!\qsetw{3cm}
[.$0$ [.$0$ [.$\bullet$ $\dots$ !\qsetw{0.1cm} $\dots$ ] !\qsetw{0.5cm} [.$\bullet$ $\dots$ !\qsetw{0.1cm} $\dots$ ] ] !\qsetw{1cm} [.$1$ [.$1$ $\dots$ !\qsetw{0.1cm} $\dots$ ] !\qsetw{0.5cm} [.$1$ $\dots$ !\qsetw{0.1cm} $\dots$ ] ] ] ]
\caption{The image under $\tau$ of a configuration $f\in \Sigma^{*}$.}\label{fig:fici}
\end{center}
\end{figure}
\end{example}

\begin{remark}\label{r:image}
Given a cellular automaton $\tau \colon A^{\Sigma^*} \to  A^{\Sigma^*}$, it immediately follows from Theorem~\ref{t:curtis} and the compactness of $A^{\Sigma^*}$, that the image $\tau(A^{\Sigma^*}) \subset A^{\Sigma^*}$ is a subshift of $A^{\Sigma^*}$.
\end{remark}

\section{Expansiveness}\label{s:expansiveness}
In this section we prove that there is no positively expansive CA on $A^{\Sigma^*}$ if $k = \vert \Sigma \vert \geq2$.
For this, we follow the proof given by Shereshevsky~\cite{Shereshevsky93} (see also~\cite{Zinoviadis10}) for CA defined on $\mathbb{Z}^d$ with $d\geq2$.

\begin{definition}[\bf Positive expansiveness]
A CA $\tau\colon A^{\Sigma^*} \to A^{\Sigma^*}$ is \emph{positively expansive} if there exists $N \geq 1$ such that for any $f_1, f_2 \in A^{\Sigma^*}$ one has $f_1 \neq f_2 \Longrightarrow \tau^t(f_1)\vert_{\Delta_N} \neq \tau^t(f_2)\vert_{\Delta_N}$, for some $t \geq 0$.
\end{definition}

\begin{remark}
The definition of positive expansiveness can be reformulated as follows: there exists $N \geq 1$ such that for any $f_1, f_2$ one has $f_1 \neq f_2 \Longrightarrow \dist(\tau^t(f_1), \tau^t(f_2)) \geq \frac1N$, for some $t \geq 0$.
\end{remark}

\begin{definition}[\bf Entropy of a CA]Let $\tau$ be a cellular automaton. We define
$$\mathcal{P}(\tau, n, t) = \{(f\vert_{\Delta_n}, \tau(f)\vert_{\Delta_n}, \dots, \tau^{t-1}(f)\vert_{\Delta_n}) : f \in A^{\Sigma^*}\}.$$
Obviously, $\mathcal{P}(\tau, n, t) \subset \underbrace{A^{\Delta_n} \times \dots \times A^{\Delta_n}}_t$ and an immediate consequence of this fact is that
$\vert\mathcal{P}(\tau, n, t)\vert \leq \vert A\vert^{\vert\Delta_n\vert t}$.
We define $h(\tau,n) = \limsup_{t \to \infty}\frac{\log\vert\mathcal{P}(\tau, n, t)\vert}{t}$.
The \emph{entropy of $\tau$} is defined by$$h(\tau) = \lim_{n \to \infty}h(\tau,n) = \sup_{n \geq 1}h(\tau,n).$$
\end{definition}

\begin{lemma}\label{l:existence t}
Let $\tau$ be an expansive CA with constant $N$. Then there exists $t \geq 1$ such that for every $n \geq 0$ and $f_1, f_2 \in A^{\Sigma^*}$ one has
$$\tau^i(f_1)\vert_{\Delta_{N+n}} = \tau^i(f_2)\vert_{\Delta_{N+n}} \forall_{i=0}^t \Longrightarrow f_1\vert_{\Delta_{N+n+1}} = f_2\vert_{\Delta_{N+n+1}}.$$
\end{lemma}
\proof
First we prove our statement for $n=0$, that is, we prove that there exists $t \geq 1$ such that for every $f_1, f_2 \in A^{\Sigma^*}$ one has
\begin{equation}\label{n=0}
\tau^i(f_1)\vert_{\Delta_{N}} = \tau^i(f_2)\vert_{\Delta_{N}} \forall_{i=0}^t \Longrightarrow f_1\vert_{\Delta_{N+1}} = f_2\vert_{\Delta_{N+1}}.
\end{equation}
Suppose the contrary. For each $t \geq 1$ there exist $f_1^{(t)}, f_2^{(t)} \in A^{\Sigma^*}$ such that
$$\tau^i(f_1^{(t)})\vert_{\Delta_{N}} = \tau^i(f_2^{(t)})\vert_{\Delta_{N}} \forall_{i=0}^t$$
and
$$f_1^{(t)}\vert_{\Delta_{N+1}} \neq f_2^{(t)}\vert_{\Delta_{N+1}}.$$
As $A^{\Sigma^*}$ is compact, consider two suitable subsequences $(f_1^{(t_h)})_{h\in\mathbb{N}}$ and $(f_2^{(t_h)})_{h\in\mathbb{N}}$ of $(f_1^{(t)})_{t\in\mathbb{N}}$ and $(f_2^{(t)})_{t\in\mathbb{N}}$ converging to
$f_1$ and $f_2$ respectively. For $h$ large enough, we have $f_1\vert_{\Delta_{N+1}} = f_1^{(t_h)}\vert_{\Delta_{N+1}}$ and $f_2\vert_{\Delta_{N+1}} = f_2^{(t_h)}\vert_{\Delta_{N+1}}$, thus $f_1\vert_{\Delta_{N+1}} \neq f_2\vert_{\Delta_{N+1}}$. Since $\tau$ is continuous, we have that
$\tau^t(f_1^{(t_h)}) \to \tau^t(f_1)$ and $\tau^t(f_2^{(t_h)}) \to \tau^t(f_2)$, for each $t\geq0$.
Once $t$ has been fixed, it holds, for $h$ large enough, $\tau^t(f_1^{(t_h)})\vert_{\Delta_{N}} = \tau^t(f_1)\vert_{\Delta_{N}}$, $\tau^t(f_2^{(t_h)})\vert_{\Delta_{N}} = \tau^t(f_2)\vert_{\Delta_{N}}$ and $t\leq t_h$. Thus, $\tau^t(f_1^{(t_h)})\vert_{\Delta_{N}} = \tau^t(f_2^{(t_h)})\vert_{\Delta_{N}}$ and then
$\tau^t(f_1)\vert_{\Delta_{N}} = \tau^t(f_2)\vert_{\Delta_{N}}$, contradicting the expansiveness of $\tau$.
Hence~\eqref{n=0} holds for a suitable $t\geq1$.

To prove our statement in general, suppose that $t\geq1$ is as in~\eqref{n=0}. Let $n \geq 0$ be an integer and let $f_1, f_2 \in A^{\Sigma^*}$ be two configurations such that $\tau^i(f_1)\vert_{\Delta_{N+n}} = \tau^i(f_2)\vert_{\Delta_{N+n}} \forall_{i=0}^t$.
Fix $w \in \Delta_{n+1}$. For every $w' \in \Delta_N$ we have that $ww' \in \Delta_{N+n}$ and $\tau^i(f_1^w)(w') = \tau^i(f_1)(ww') = \tau^i(f_2)(ww') = \tau^i(f_2^w)(w') \forall_{i=0}^t$. That is, $\tau^i(f_1^w)\vert_{\Delta_{N}} = \tau^i(f_2^w)\vert_{\Delta_{N}} \forall_{i=0}^t$. By~\eqref{n=0}, we have that $f_1^w\vert_{\Delta_{N+1}} = f_2^w\vert_{\Delta_{N+1}}$. The word $w\in \Delta_{n+1}$ having been chosen arbitrarily, we have that $f_1\vert_{\Delta_{N+n+1}} = f_2\vert_{\Delta_{N+n+1}}$
\endproof

\begin{lemma}\label{l:nt}
Let $\tau$ be an expansive CA with constant $N$ and $t \geq 1$ as in Lemma~\ref{l:existence t}. Then for every $n \geq 1$ and $f_1, f_2 \in A^{\Sigma^*}$ one has
$$\tau^i(f_1)\vert_{\Delta_N} = \tau^i(f_2)\vert_{\Delta_N} \forall_{i=0}^{nt} \Longrightarrow f_1\vert_{\Delta_{N+n}} = f_2\vert_{\Delta_{N+n}}.$$
\end{lemma}
\proof
By Lemma~\ref{l:existence t}, from $\tau^i(f_1)\vert_{\Delta_N} = \tau^i(f_2)\vert_{\Delta_N} \forall_{i=0}^{nt}$, we have $\tau^i(f_1)\vert_{\Delta_{N+1}} = \tau^i(f_2)\vert_{\Delta_{N+1}} \forall_{i=0}^{(n-1)t}$. Again, applying Lemma~\ref{l:existence t}, we have
$\tau^i(f_1)\vert_{\Delta_{N+2}} = \tau^i(f_2)\vert_{\Delta_{N+2}} \forall_{i=0}^{(n-2)t}$. By iterating this argument, the claim follows.
\endproof

A consequence of Lemma~\ref{l:nt} is the following.

\begin{corollary}\label{c:corexp} Let $\tau$ be an expansive CA with constant $N$ and $t \geq 1$ as in Lemma~\ref{l:existence t}. Then for every $n\geq 1$
$$\vert A \vert^{\vert\Delta_{N+n}\vert} \leq \vert\mathcal{P}(\tau, N, nt+1)\vert.$$
\end{corollary}

\begin{theorem}\label{t:nonexp}For $\vert\Sigma\vert\geq2$, there do not exist expansive CA on $A^{\Sigma^*}$.
\end{theorem}

\proof
Suppose that $\tau$ is expansive with constant $N$, and let $t \geq 1$ be as in Lemma~\ref{l:existence t}. By Corollary~\ref{c:corexp}, we have that
$\vert A \vert^{\vert\Delta_{N+n}\vert} \leq \vert\mathcal{P}(\tau, N, nt+1)\vert$ for every $n\geq 1$. Moreover, by definition,  we have $\vert\mathcal{P}(\tau, n, t)\vert \leq \vert A\vert^{\vert\Delta_n\vert t}$ for each $n \geq 1$ and $t \geq 0$. In particular $\vert\mathcal{P}(\tau, N, nt+1)\vert \leq \vert A\vert^{\vert\Delta_N\vert(nt+1)}$. This implies
$\vert A \vert^{\vert\Delta_{N+n}\vert} \leq \vert A\vert^{\vert\Delta_N\vert (nt+1)}$, that is,  $\frac{k^{N+n}-1}{k-1} \leq \frac{(k^N-1)}{k-1}(nt+1)$ (where $k = \vert\Sigma\vert$). This is a contradiction because $n$ can be chosen arbitrarily large.
\endproof

%
\section{Permutivity, preinjectivity and surjectivity}\label{s:permutivity}

In what follows,  $\tau\colon A^{\Sigma^*} \to A^{\Sigma^*}$ denotes a CA defined by a local map $\mu\colon A^{\Delta_{r+1}} \to A$. We always suppose that $r\geq1$.

Fix $n \geq 1$. For a pattern $p \colon \Delta_n \setminus \{\varepsilon\} \to A$ and a letter $a \in A$, we denote by $a \triangleleft p$
the block on $A^{\Delta_n}$ coinciding with $p$ on $\Delta_n \setminus \{\varepsilon\}$ and sending $\varepsilon$ to $a$.

\begin{definition}[\bf Permutivity]\label{d:permutivity} A CA $\tau$ of radius $r\geq1$ is \emph{permutive} if for every pattern $p \colon \Delta_{r+1} \setminus \{\varepsilon\} \to A$ and every  letter $a \in A$, there exists a unique $a' \in A$ such that $\mu(a' \triangleleft p) = a$.
\end{definition}

In other words, $\tau$ is permutive if for every $p \in A^{\Delta_{r+1} \setminus \{\varepsilon\}}$ the function $\mu((\cdot) \triangleleft p) : A \to A$ is a permutation of the alphabet $A$.

\begin{proposition}\label{p:permutive -> surjective} Permutive CA are surjective.
\end{proposition}
\proof
Let $f \in A^{\Sigma^*}$ be a configuration. For every $n \geq 1$ there exists a configuration $f_n$ such that $\tau(f_n)\vert_{\Delta_n} = f\vert_{\Delta_n}$. For this, it is sufficient to arbitrarily fix the levels from the $(n+1)$-th to the $(n+r)$-th in the configuration $f_n$ and fill it in backward by permutivity.

This implies that $\lim_{n \to \infty}\tau(f_n) = f$.
Since $A^{\Sigma^*}$ is compact, there exists a subsequence $(f_{n_h})_{h\in\mathbb{N}}$ of $(f_n)_{n\in\mathbb{N}}$ converging to some $g \in A^{\Sigma^*}$. Thus, $\lim_{h \to \infty}\tau(f_{n_h}) = \tau(g)$ and then $\tau(g) = f$.
\endproof

\begin{definition}[\bf Diamond]\label{d:diamond}
Let $\tau$ be a CA and let $p \in A^{\Delta_r}$ be a block. A \emph{diamond of $\tau$ (based on $p$)} is a pair of different blocks having the same support of size $n > 2r+2$ coinciding with $p$ on $\Delta_r$ and on $v\Delta_r$ for each vertex $v$ of length $n-r$, and having the same image under $\mu$.
\end{definition}

\begin{remark}\label{r:diamond}
Let $\tau$ be a CA and $p \in A^{\Delta_r}$ a block. If $p_1, p_2 \in A^{\Delta_n}$ are two blocks (with $n > 2r+2$) coinciding with $p$ on $\Delta_r$ and on $v\Delta_r$ for each vertex $v$ of length $n-r$, we can construct two configurations $f_1, f_2 \in A^{\Sigma^*}$ such that $f_i$ repeats $p_i$, $i = 1,2$, infinitely many times. Note that $f_1 = f_2$ if and only if $p_1 = p_2$, and $\tau(f_1) = \tau(f_2)$ if and only if $\mu(p_1) = \mu(p_2)$.
\end{remark}

\begin{definition}[\bf Preinjectivity]\label{d:preinjectivity} A CA $\tau$ is \emph{preinjective} if it has no diamonds.
\end{definition}

\begin{proposition}\label{p:permutive -> preinjective} Permutive CA are preinjective.
\end{proposition}
\proof
Let $\tau$ be a permutive CA. For $i=1,2$, let $p_i$ and $f_i$ be as in Remark~\ref{r:diamond}. Suppose that $f_1 \neq f_2$ are different configurations. Then there exists $v \in \Sigma^*$ such that $f_1(v) \neq f_2(v)$ and $f_1\vert_{v\Delta_{r+1} \setminus \{v\}}=f_2\vert_{v\Delta_{r+1} \setminus \{v\}}$. Equivalently, we have that
$f_1^v(\varepsilon) \neq f_2^v(\varepsilon)$ and $f_1^v\vert_{\Delta_{r+1} \setminus \{\varepsilon\}}=f_2^v\vert_{\Delta_{r+1} \setminus \{\varepsilon\}}$.
Since $\tau$ is permutive, we have $\mu((f_1^v)\vert_{\Delta_{r+1}}) \neq \mu((f_2^v)\vert_{\Delta_{r+1}})$. Hence
$$
\tau(f_1)(v) = \mu((f_1^v)\vert_{\Delta_{r+1}}) \neq \mu((f_2^v)\vert_{\Delta_{r+1}}) = \tau(f_2)(v),
$$
and thus $\tau(f_1) \neq \tau(f_2)$ and $\mu(p_1) \neq \mu(p_2)$. This implies that $\tau$ has no diamonds.
%
\endproof

\subsection{An example of a surjective CA which is not preinjective}\label{ss:non moore}

\begin{figure}[h!]
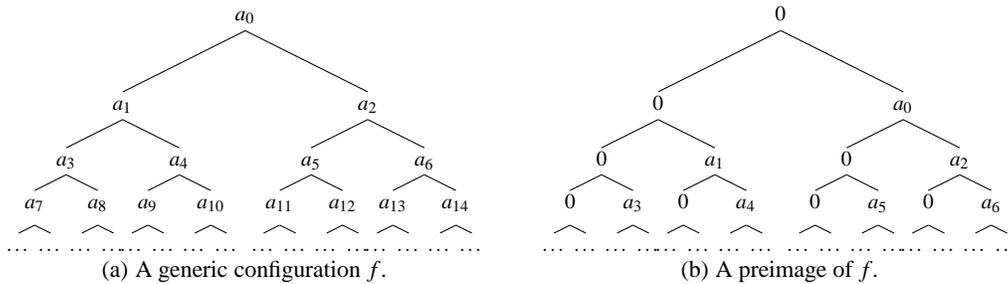

\centering
\subfigure[A generic configuration $f$.]{
\qtreecenterfalse
\scriptsize
\Tree [.$a_0$
[.$a_1$ [.$a_3$ [.$a_7$ $\dots$ !\qsetw{0.1cm} $\dots$ ] !\qsetw{0.5cm} [.$a_8$ $\dots$ !\qsetw{0.1cm} $\dots$ ] ] !\qsetw{1cm} [.$a_4$ [.$a_9$ $\dots$ !\qsetw{0.1cm} $\dots$ ] !\qsetw{0.5cm} [.$a_{10}$ $\dots$ !\qsetw{0.1cm} $\dots$ ] ] ]
!\qsetw{3cm}
[.$a_2$ [.$a_5$ [.$a_{11}$ $\dots$ !\qsetw{0.1cm} $\dots$ ] !\qsetw{0.5cm} [.$a_{12}$ $\dots$ !\qsetw{0.1cm} $\dots$ ] ] !\qsetw{1cm} [.$a_6$ [.$a_{13}$ $\dots$ !\qsetw{0.1cm} $\dots$ ] !\qsetw{0.5cm} [.$a_{14}$ $\dots$ !\qsetw{0.1cm} $\dots$ ] ] ] ]\label{f: generic configuration}
}\quad\quad
\subfigure[A preimage of $f$.]{
\qtreecenterfalse
\scriptsize
\Tree [.$0$
[.$0$ [.$0$ [.$0$ $\dots$ !\qsetw{0.1cm} $\dots$ ] !\qsetw{0.5cm} [.$a_3$ $\dots$ !\qsetw{0.1cm} $\dots$ ] ] !\qsetw{1cm} [.$a_1$ [.$0$ $\dots$ !\qsetw{0.1cm} $\dots$ ] !\qsetw{0.5cm} [.$a_4$ $\dots$ !\qsetw{0.1cm} $\dots$ ] ] ]
!\qsetw{3cm}
[.$a_0$ [.$0$ [.$0$ $\dots$ !\qsetw{0.1cm} $\dots$ ] !\qsetw{0.5cm} [.$a_5$ $\dots$ !\qsetw{0.1cm} $\dots$ ] ] !\qsetw{1cm} [.$a_2$ [.$0$ $\dots$ !\qsetw{0.1cm} $\dots$ ] !\qsetw{0.5cm} [.$a_6$ $\dots$ !\qsetw{0.1cm} $\dots$ ] ] ] ]\label{f: preimage}
}
\caption{An example of surjective CA.}
\label{fig:surjective}
\end{figure}

Let $A = \{0,1\}$ and $r=1$. Consider the CA $\tau$ whose local defining map $\mu \colon A^{\Delta_2} \to A$ defined by
\[
\begin{split}
\mu(p) & =
p(0) + p(1) \mod 2.
\end{split}
\]
Figure~\ref{f: generic configuration} displays a generic configuration $f$, and Figure~\ref{f: preimage} a preimage of $f$.

\begin{figure}[!h]
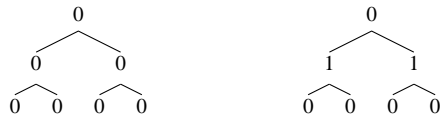

\begin{center}
\qtreecenterfalse
\scriptsize
\Tree [.$0$ [.$0$ [.$0$ ] [.$0$ ] ] [.$0$ [.$0$ ] [.$0$ ] ] ]
\hspace{2cm}
\Tree [.$0$ [.$1$ [.$0$ ] [.$0$ ] ] [.$1$ [.$0$ ] [.$0$ ] ] ]
\caption{A diamond of $\tau$.}\label{f: non preinjective}
\end{center}
\end{figure}

The CA $\tau$ is not preinjective: Consider the patterns $p_1, p_2 \in A^{\Delta_3}$ such that $p_1(v) = 0 = p_2(v)$ for each $v \notin \{0,1\}$ and $f_1(0) = f_1(1) = 0 = 1-f_2(0) = 1-f_2(1)$ (see Figure~\ref{f: non preinjective}). Clearly, we have $\mu(p_1) = \mu(p_2)$.

\subsection{The Garden of Eden theorem}
As it is well-known, preinjectivity and surjectivity are equivalent for CA over $A^\mathbb{N}$ or even more generally over $A^{\mathbb{Z}^d}$. This is the statement of the \emph{Garden of Eden theorem} proved by Moore~\cite{Moore62} and Myhill~\cite{Myhill63}.

In this section we focus on the
 the implication ``preinjectivity $\Longrightarrow$ surjectivity''. In the case of $\mathbb{Z}$ it can be proved for a wide class of subshifts using arguments involving the entropy (see~\cite{Fiorenzi00}). In~\cite[Proposition 5.26]{Kurka03}, K\r{u}rka provides a direct proof using the following fact: if $\tau$ is not surjective, then there exists a word $v \in A^n$ such that the preimage set $\mu^{-1}(v)$ has cardinality $\xi > \vert A\vert^r$ (where $r$ is the radius of $\tau$). With this latter result at hand, fix $w \in A^{n+r}$ such that $w \in \mu^{-1}(v)$. For each $m\geq1$ define the sets
$$\M_m = \{w\ w_1\ \cdots w_m\ w : w_i \in \mu^{-1}(v)\} \quad \textup{ and } \quad
\M_m' = \{v\ v_1\ v\ \cdots v\ v_{m+1}\ v : v_i \in A^r\}.
$$
One has $\mu(\M_m) \subset \M_m'$, $\vert \M_m \vert = \xi^m$ and $\vert \M_m' \vert = A^{r(m+1)}$. Since $\xi > \vert A \vert^r$,  one has
$\xi^m > \vert A\vert^{r(m+1)}$ for $m$ large enough. This implies that distinct elements of $\M_m$ have the same image, i.e., there exists at least one diamond.

We prove now that the same arguments apply in our general case. In particular the following proposition is an easy generalization of a theorem by K\r{u}rka~\cite[Theorem 5.21]{Kurka03} (see also~\cite{Hedlund69}).

\begin{proposition}\label{p:card=}
Let $\tau$ be a CA of radius $r\geq1$. If for every $n\geq1$ and every block $q \in A^{\Delta_n}$ the preimage set $\mu^{-1}(q)$ has cardinality $\vert A\vert^{\vert\Delta_{n+r}\vert-\vert\Delta_n\vert}$, then $\tau$ is surjective.
\end{proposition}
\proof
Let $f \in A^{\Sigma^*}$ be a configuration. For $n\geq1$ set
$$
X_n = \{g \in A^{\Sigma^*} : g\vert_{\Delta_{n+r}} \stackrel{\mu}{\longrightarrow} f\vert_{\Delta_n}\}.
$$
Then $X_n = \tau^{-1}(\C(f\vert_{\Delta_n}))$, that is the preimage of the cylinder determined by the block $f\vert_{\Delta_n}$. By hypothesis, $X_n$ is nonempty and it is closed since $\tau$ is continuous. Moreover $X_{n+1} \subset X_n$ for every $n \geq1$. By compactness, there exists $g \in \bigcap_{n\geq1}X_n$ and obviously $\tau(g)=f$. Thus, $\tau$ is surjective.
\endproof

\begin{corollary}\label{c:nonsur}
Let $\tau$ be a CA of radius $r\geq1$. If $\tau$ is not surjective then there exist an integer $n\geq1$ and a block $q \in A^{\Delta_n}$ such that $\vert \mu^{-1}(q) \vert > \vert A\vert^{\vert\Delta_{n+r}\vert-\vert\Delta_n\vert}$.
\end{corollary}
\proof
Suppose that $\tau$ is not surjective.
By Proposition~\ref{p:card=}, we have that there exist an integer $n\geq1$ and a block $q \in A^{\Delta_n}$ such that $\vert \mu^{-1}(q) \vert \neq \vert A\vert^{\vert\Delta_{n+r}\vert-\vert\Delta_n\vert}$. Notice that, once $n$ has been fixed, $\bigcup_{q\in A^{\Delta_n}} \mu^{-1}(q) = A^{\Delta_{n+r}}$ and then
$$
\sum_{q\in A^{\Delta_n}} \vert\mu^{-1}(q)\vert = \vert A\vert^{\vert\Delta_{n+r}\vert}.
$$
Thus the mean number of preimages of a block $q \in A^{\Delta_n}$ is $\vert A\vert^{\vert\Delta_{n+r}\vert-\vert\Delta_n\vert}$ and if $\vert \mu^{-1}(q) \vert < \vert A\vert^{\vert\Delta_{n+r}\vert-\vert\Delta_n\vert}$, there must exist $q' \in A^{\Delta_n}$ such that
$\vert \mu^{-1}(q') \vert > \vert A\vert^{\vert\Delta_{n+r}\vert-\vert\Delta_n\vert}$.
\endproof

\begin{theorem}\label{t:preinjective -> surjective}
Preinjective CA are surjective.
\end{theorem}
\proof
Apply the argument of K\r{u}rka illustrated above. Suppose that a CA $\tau$ is not surjective. By virtue of Corollary~\ref{c:nonsur}, there exists a pattern $q \in A^{\Delta_n}$ such that the preimage set $\mu^{-1}(q)$ has cardinality $\xi > \vert A \vert^{\vert\Delta_{n+r}\vert-\vert\Delta_n\vert} = \vert A \vert^\frac{k^{n+r}-k^n}{k-1}$. Fix $p \in A^{\Delta_{n+r}}$ such that $p \in \mu^{-1}(q)$. The sets $\M_m$ and $\M_m'$ are defined analogously and one has $\mu(\M_m) \subset \M_m'$. In this case we have $\vert \M_m \vert = \xi^{\frac{(k^{n+r})^{m+1}-1}{k^{n+r}-1}-1}$ and $\vert \M_m' \vert = \vert A\vert^{\frac{(k^{n+r})^{m+1}-1}{k^{n+r}-1}\frac{k^{n+r}-k^n}{k-1}} = \left(\vert A\vert^\frac{k^{n+r}-k^n}{k-1}\right)^{\frac{(k^{n+r})^{m+1}-1}{k^{n+r}-1}}$. Again, since $\xi > \vert A \vert^\frac{k^{n+r}-k^n}{k-1}$,  one has
$\xi^{\frac{(k^{n+r})^{m+1}-1}{k^{n+r}-1}-1} > \left(\vert A\vert^\frac{k^{n+r}-k^n}{k-1}\right)^{\frac{(k^{n+r})^{m+1}-1}{k^{n+r}-1}}$ for $m$ large enough. This implies that distinct elements of $\M_m$ have the same image, i.e., there exists at least one diamond.
\endproof

\begin{remark}
Notice that Proposition~\ref{p:permutive -> surjective} is also a consequence of Proposition~\ref{p:permutive -> preinjective} and 
Theorem~\ref{t:preinjective -> surjective}.
\end{remark}

\section{Closingness}\label{s:closingness}
Right-closingness is a well-known notion in the one-dimensional case and positively expansive CA share this property~\cite{Kurka97} (see also \cite[Lemma 6.5]{BoyleKitchens99}).
In this section we generalize this notion to CA on the full tree shift.

Let $p \in A^{\Delta_n}$ and $q \in A^{\Delta_m}$ be two patterns of a CA $\tau$, with $n,m \geq 1$. The notation $p \stackrel{\mu}{\longrightarrow} q$ means that there exists a configuration $f \in A^{\Sigma^*}$ such that $f\vert_{\Delta_n} = p$ and $\tau(f)\vert_{\Delta_m} = q$.

Fix $n,m \geq 1$. Given a block $p \in A^{\Delta_n}$ and a sequence of blocks $(p_1, \dots, p_{k^n}) \in A^{\Delta_m} \times \dots \times A^{\Delta_m}$, we denote by $p \triangleleft (p_1, \dots, p_{k^n})$ the pattern of $A^{\Delta_{n+m}}$ coinciding with $p$ on $\Delta_n$ and with $p_i$ on $v_i\Delta_r$ for each $i = 1, \dots, k^n$, where $v_i$ is the $i$-th word over $\Sigma$ of length $n$ (in the lexicographical order).

\begin{definition}[\bf Right-closingness]\label{d:right-closingness} A CA $\tau$ is \emph{right-closing} if there exists $N \in \mathbb{N}$ such that if
$p \stackrel{\mu}{\longrightarrow} q$ with $\supp(p) = \Delta_{r}$ and $\supp(q) = \Delta_{rN}$, then there exists a unique sequence $(p_1, \dots, p_{k^r}) \in (A^{\Delta_r})^{k^r}$ such that
for any configuration $f$
$$f\vert_{\Delta_r} = p \textup{ and } \tau(f)\vert_{\Delta_{rN}} = q \Longrightarrow f\vert_{v_i\Delta_r} = p_i \quad \forall_{i=1}^{k^r},$$
where $v_i$ is the $i$-th word over $\Sigma$ of length $r$ in the lexicographical order.
\end{definition}

\begin{proposition}\label{p:right-closing -> preinjective} Right-closing CA are preinjective.
\end{proposition}
\proof
For $i=1,2$, let $p_i$ and $f_i$ be as in Remark~\ref{r:diamond} and suppose $\tau(f_1) = \tau(f_2)$. Right-closingness implies that
$f_1\vert_{v\Delta_r} = f_2\vert_{v\Delta_r}$ for each $v \in \Sigma^r$. Again, we have that $f_1\vert_{u v \Delta_r} = f_2\vert_{u v \Delta_r}$ for each $u, v \in \Sigma^r$. So we have that $f_1$ and $f_2$ agree on triangles of increasing size. This implies $f_1 = f_2$ and then $p_1 = p_2$, and thus $\tau$ has no diamonds.
\endproof

Proposition~\ref{p:right-closing -> preinjective} and Theorem~\ref{t:preinjective -> surjective} imply the following result.

\begin{corollary}\label{p:right-closing -> surjective}
Right-closing CA are surjective.
\end{corollary}

\subsection{The one-dimensional case}
If $\tau$ is a CA on $A^{\mathbb{Z}}$, there exists a more natural definition of closingness that uses the linear structure of $\mathbb{Z}$. We illustrate it below. For more details see~\cite{BoyleKitchens99}, \cite{Kurka03} and~\cite{BlanchardMaass97}.

\begin{definition}[\bf Left-asymptoticity] Two configurations $f_1, f_2 \in A^\mathbb{Z}$ are \emph{left-asymptotic} if $f_1\vert_{]-\infty, n]} = f_2\vert_{]-\infty, n]}$ for some $n \in \mathbb{Z}$.
\end{definition}

\begin{definition}[\bf Right-closingness in $\mathbb{Z}$ and $\mathbb{N}$] A CA $\tau \colon A^\mathbb{Z} \to A^\mathbb{Z}$ is \emph{right-closing} if any two different left-asymptotic configurations $f_1, f_2 \in A^\mathbb{Z}$ verify $\tau(f_1) \neq \tau(f_2)$. A CA $\tau \colon A^\mathbb{N} \to A^\mathbb{N}$ is \emph{right-closing} if the natural extension $\bar \tau \colon A^\mathbb{Z} \to A^\mathbb{Z}$ of $\tau$ (i.e., $\bar \tau$ is defined by the same local map as $\tau$), is right-closing.
\end{definition}

In \cite[Section 3.1]{BlanchardMaass97}, Blanchard and Maass give the following ``finitary'' definition of right-closingness in $\mathbb{N}$. We extended this definition to CA on trees in Definition~\ref{d:right-closingness}.

\begin{definition} A CA $\tau \colon A^\mathbb{N} \to A^\mathbb{N}$ is right-closing if there exists $n\in\mathbb{N}$ such that if
$v \stackrel{\mu}{\longrightarrow} w_1 \cdots w_N$ with $v, w_i \in A^r$, $i=1, \dots, N$, then there exists a unique $v' \in A^r$ such that for any configuration $f\in A^\mathbb{N}$ one has
$$f\vert_{[0,r-1]} = v \textup{ and } \tau(f)\vert_{[0,rN-1]} = w_1 \cdots w_N \Longrightarrow f\vert_{[r,2r-1]} = v'.$$
\end{definition}

\section{Openness}\label{s:openness}

\begin{definition}[\bf Openness]\label{d:openness}
A CA $\tau$ is \emph{open} if the image of an open set is an open set.
\end{definition}
\begin{remark}
A CA is open if and only if the image of a cylinder is an union of cylinders.
\end{remark}
\begin{proposition}\label{p:open -> surjective} Open CA are surjective.
\end{proposition}
\proof
Suppose that $\tau$ is open. Fix a letter $a\in A$ and let $\C(a)$ be the cylinder determined by $a$. By assumption, there exists a block $p$ such that $\C(p) \subset \tau(\C(a))$, where $\C(p)$ is the cylinder determined by $p$. In particular we have that $a \stackrel{\mu}{\longrightarrow} p$.
Suppose that $\supp(p) = \Delta_m$ and define $v = 0^m \in \Sigma^m$. Consider a configuration $f \in A^{\Sigma^*}$. We want to prove that $f$ has a preimage. Fix a configuration $f_1$ coinciding with $p$ on $\Delta_m$ and such that $f_1^v = f$. By definition, we have that $f_1 \in \C(p)$ and then there exists $g \in \C(a)$ such that $\tau(g) = f_1$. Thus, $\tau(g^v) = \tau(g)^v = f_1^v = f$.\endproof

\begin{remark}
Proposition~\ref{p:open -> surjective} can be proved by means of a topological argument. Indeed, if $\tau$ is open, we have that $\tau(A^{\Sigma^*})$ is open. This implies that $\tau(A^{\Sigma^*}) = A^{\Sigma^*}$, because $\tau(A^{\Sigma^*})$ is a tree shift and the only open tree shifts are $\emptyset$ and $A^{\Sigma^*}$.
To prove this claim, consider a susbshift $X$ and a configuration $f\in X$. Fix a forbidden block $p$ for $X$. We define a sequence $(f_n)_{n\in \mathbb{N}} \in A^{\Sigma^*}$ in such a way that $f_n$ and $f$ agree on $\Delta_n$. Moreover, we impose $f_n^v = p$ for some $v \notin \Delta_n$. In this way, we have $\lim_{n \to \infty}f_n = f$ and $p$ appears in each $f_n$ (hence $f_n \in A^{\Sigma^*} \setminus X$). If $X$ were open, we would have that $f\in A^{\Sigma^*}\setminus X$, which is a contradiction. We do not get a contradiction when: (1) there do not exist configurations $f
\in X$ (i.e., $X$ is empty), or (2) there does not exist a forbidden block $p$ for $X$ (i.e., $X$ is the full shift). In the case of bidimensional CA, the same argument is used in~\cite[Proposition 4]{DennunzioFormenti08}.

\end{remark}

In \cite{BlanchardMaass97}, Blanchard and Maass prove the following result for CA on $A^\mathbb{N}$. In Proposition~\ref{p:right-closing -> open}, we give a generalization of their proof in the case of CA on $A^{\Sigma^*}$.

\begin{proposition}\label{p:BM}\cite[Proposition 3.2]{BlanchardMaass97} Right-closing CA on $A^\mathbb{N}$ are open.
\end{proposition}

We want to point out that in $A^\mathbb{N}$ it is possible to recode a CA of radius $r$ with an equivalent CA of radius~$1$. This recoding is more complicated in $A^{\Sigma^*}$. This explains the additional hypotheses in the following proposition.

\begin{figure}[h!]
\centering
\begin{picture}(0,40)(0,-40)
\gasset{Nw=0,Nh=0,Nframe=n,ATnb=0,AHnb=0}

\scriptsize
\node(0)(0,0){$\not \stackrel{\phantom{a}\mu}{\longrightarrow}$}
\node[Nw=4,Nh=4](s0)(-20,0){$a'$}
\node(s1)(-29,-22){}
\node(s2)(-11,-22){}
\drawedge(s0,s1){}
\drawedge(s1,s2){}
\drawedge(s2,s0){}
\node(s3)(-20,-13){$p$}
\node(s4)(-20,-25){$a\quad\quad\ \ \dots\quad\quad\ \ a$}
\node(s5)(-31,-27){}
\node(s6)(-9,-27){}
\drawedge(s5,s6){}
\drawedge(s5,s1){}
\drawedge(s6,s2){}
\node(p1)(-36,-13.5){$p'$}
\drawcurve[dash={1}0](-22,0)(-34,-13.5)(-32,-27)

\node(d1)(15,-12){}
\node(d2)(25,-12){}
\drawedge(d1,d2){}
\node(d3)(20,-8){$q$}
\node(d4)(20,-15){$b_1 \dots\ b_{2^N}$}
\node(d6)(13,-17){}
\node(d7)(27,-17){}
\drawedge(d6,d7){}
\drawpolygon[Nframe=y](11,-22)(20,0)(29,-22)
\node(d5)(20,-20){$\dots$}

\node(ddd)(20,-30){$\dots$}

\node(dd1)(6,-34){}
\node(dd2)(16,-34){}
\drawpolygon[Nframe=y](6,-34)(11,-22)(16,-34)(18,-39)(4,-39)
\drawedge(dd1,dd2){}
\node(dd3)(11,-30){$q$}
\node(dd4)(11,-37){$b_1 \dots\ b_{2^N}$}

\node(ddd1)(24,-34){}
\node(ddd2)(34,-34){}
\drawpolygon[Nframe=y](24,-34)(29,-22)(34,-34)(36,-39)(22,-39)
\drawedge(ddd1,ddd2){}
\node(ddd3)(29,-30){$q$}
\node(ddd4)(29,-37){$b_1 \dots\ b_{2^N}$}

\node(q1)(39,-19,5){$q'$}
\drawcurve[dash={1}0](22,0)(37,-17.5)(35,-34)

\node(q2)(44,-19,5){$q''$}
\drawcurve[dash={1}0](24,0)(42,-19.5)(40,-39)

\end{picture}
\caption{Illustration of the construction in the proof of Proposition~\ref{p:right-closing -> open}.}
\label{fig:triangles}
\end{figure}
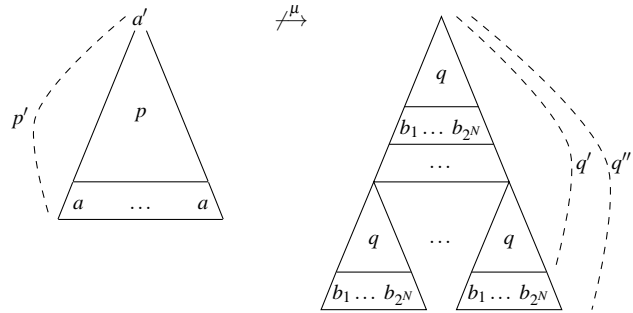

\begin{proposition}\label{p:right-closing -> open} Right-closing CA of radius $r=1$ are open.
\end{proposition}
\proof
For simplicity of the notation, we suppose $k = \vert \Sigma \vert = 2$, but the proof can be easily adapted for any $k\geq 2$.
Let $N$ be as in Definition~\ref{d:right-closingness}.  By hypothesis, we have that if
$a \stackrel{\mu}{\longrightarrow} q$, with $a \in A$ and $\supp(q) = \Delta_{N}$, then there exists a unique pair $(a_1, a_2) \in A^2$ such that
for any configuration $f$
$$f(\varepsilon) = a \textup{ and } \tau(f)\vert_{\Delta_{N}} = q \Longrightarrow f(0) = a_1 \textup{ and } f(1) = a_2.$$
We want to prove that for every tuple $(b_1, \dots, b_{2^N}) \in A^{2^N}$ there is a unique pair $(a_1, a_2) \in A^2$ such that $a \triangleleft (a_1, a_2) \stackrel{\mu}{\longrightarrow} q \triangleleft (b_1, \dots, b_{2^N})$.
Suppose the contrary. Then there exists $a \stackrel{\mu}{\longrightarrow} q$ but $a \not \stackrel{\phantom{a}\mu}{\longrightarrow} q\triangleleft (b_1, \dots, b_{2^N})$.
Since $\tau$ is surjective, we have that $a' \stackrel{\mu}{\longrightarrow} q\triangleleft (b_1, \dots, b_{2^N})$ for some letter $a' \neq a$.
Hence, there exists a pattern $p \colon \Delta_{N+2} \setminus \{\varepsilon\} \to A$ such that $a' \triangleleft p \stackrel{\mu}{\longrightarrow} q\triangleleft (b_1, \dots, b_{2^N})$.
Consider the pattern $p' \colon \Delta_{N+3} \to A$ extending $a' \triangleleft p$ in such a way that $p'(v) = a$ for each $v\in \Sigma^{N+2}$. Clearly,
$p' \stackrel{\mu}{\longrightarrow} q'$, where $q'$ is some extension of $q \triangleleft (b_1, \dots, b_{2^N})$ to $\Delta_{2N+2}$ such that $q'$ coincides with $q$ on $v\Delta_N$, for each $v \in \Sigma^{N+3}$. Consider now the block $q''$ defined as an extension to $\Delta_{2N+3}$ of $q'$ such that $q''$ coincides with $2^{N+2}$ copies of $(b_1, \dots, b_{2^N})$ on $\Sigma^{2N+2}$. By right-closingness, we have that
$a' \not \stackrel{\phantom{a}\mu}{\longrightarrow} q''$ (see Figure~\ref{fig:triangles}), and again, since $\tau$ is surjective, there exists a letter $a''$ such that $a'' \neq a'$, $a'' \neq a$ and $a'' \stackrel{\mu}{\longrightarrow} q''$. In this way, since $A$ is finite, we can eventually find a block of $A^{\Sigma^*}$ having no preimage, in contradiction with the surjectivity of $\tau$.
Hence the claim is true.

Suppose $a \stackrel{\mu}{\longrightarrow} q$. We want to prove that $\C(q) \subset \tau(\C(a))$. For any configuration $f \in \C(q)$, there exists a pair $(a_1,a_2)$ such that $a \triangleleft (a_1,a_2) \stackrel{\mu}{\longrightarrow} q \triangleleft f\vert_{\Sigma^N}$. Again, there exists a pair $(a_i',a_i'')$ such that $a_i \triangleleft (a_i',a_i'') \stackrel{\mu}{\longrightarrow} q^{i-1} \triangleleft f\vert_{\Sigma^{N+1} \cap (i-1)\Sigma^N}$, for $i=1,2$.
Hence,  $a \triangleleft (a_1,a_2) \triangleleft (a_1',a_1'',a_2',a_2'') \stackrel{\mu}{\longrightarrow} q \triangleleft f\vert_{\Sigma^N\cup\Sigma^{N+1}}$.
In this way, we can recursively construct a preimage of $f$ in $\C(a)$.
\endproof

\begin{remark}
Corollary~\ref{p:right-closing -> surjective} is also a consequence of Proposition~\ref{p:right-closing -> open} (if generalized to any radius), and Proposition~\ref{p:open -> surjective}.
It can also be obtained by the lifting (up and down) operation described in~\cite{AcerbiDennunzioFormenti09}.
\end{remark}

\section{Conclusion and possible developments}

This paper is a first attempt to study topological properties of CA on the full tree shift $A^{\Sigma^*}$. We showed that there do not exist positively expansive CA if $\vert\Sigma\vert\geq2$ (Theorem~\ref{t:nonexp}). In the case of CA on $A^{\mathbb{Z}^{d}}$, it is well known that positively expansive CA exist if and only if $d=1$.

In Section~\ref{s:permutivity}, we gave a definition of permutivity which is a natural generalization of the notion of left-permutivity for one-dimensional CA. The symmetric notion of right-permutivity is not naturally generalizable in our setting. It is easy to prove that in the one-dimensional case right-permutive CA are right-closing. In Section~\ref{s:closingness}, we defined right-closingness for CA on $A^{\Sigma^*}$ and we wonder whether there is a good definition of right-permutivity which still implies right-closingness.
In the case of bidimensional CA, some interesting constructions are given by Dennunzio and Formenti in~\cite{DennunzioFormenti08}.

For $d$-dimensional CA, preinjectivity is equivalent to surjectivity. We showed in Section~\ref{ss:non moore} that there exist CA on the full (binary) tree shift that are surjective but not preinjective. In Theorem~\ref{t:preinjective -> surjective} we prove that preinjective CA are surjective. We proved that permutivity implies surjectivity and preinjectivity (Propositions~\ref{p:permutive -> surjective} and~\ref{p:permutive -> preinjective}, respectively).
In Proposition~\ref{p:right-closing -> preinjective} we also proved that right-closingness implies preinjectivity.

In Section~\ref{s:openness} we considered open CA. We proved that openness implies surjectivity (Proposition \ref{p:open -> surjective}). In Proposition~\ref{p:BM}, we showed that right-closing CA of radius one are open. We believe that the result is generalizable to any radius. 

Other properties we are working on for CA on tree shifts are the transitivity, the mixing property, and the density of the periodic orbits. For example, in the one-dimensional case, it is known that left-permutive CA are mixing~\cite{CattaneoDennunzioMargara02}, and right-closingness implies the density of the periodic orbits~\cite{BoyleKitchens99}.

\bibliographystyle{eptcs}
\bibliography{biblio}
\end{document}